\documentclass[aps,twocolumn,float,prd,psfig,amsmath,booktabs]{revtex4}
\usepackage[dvipdfmx]{graphicx}
\usepackage{amsmath,amssymb,natbib,siunitx,booktabs,url}
\usepackage{color}

\newcommand{\add}[1]{\textcolor{black}{#1}}

\newcommand{\lsim}{\lesssim}
\newcommand{\gsim}{\gtrsim}

\newcommand{\CC}{{K}}
\newcommand{\vect}[1]{\mbox{\boldmath${#1}$}}
\newcommand{\lmk}{\left(}
\newcommand{\rmk}{\right)}
\newcommand{\lnk}{\left\{ }
\newcommand{\rnk}{\right\} }
\newcommand{\lkk}{\left[}
\newcommand{\rkk}{\right]}
\newcommand{\lla}{\left\langle}
\newcommand{\p}{\partial}
\newcommand{\rra}{\right\rangle}
\newcommand{\so}{M_\odot}
\newcommand{\mch}{{\cal M}}

\newcommand{\bea}{{\begin{eqnarray}}}
\newcommand{\eea}{{\end{eqnarray}}}

\begin{document}

\title{Perturbative Frequency Expansion for Nearly Monochromatic Binary Black Holes Detectable with LISA }

\author{Naoki Seto }
%%%%%%%%%%%%%%%%%%%%%%%%%%%%%%%%%%%%%%%%%%%%%%%%%%%%%%%%%%%%%%%%%%%%%%%%%%
\affiliation{Department of Physics, Kyoto University, 
Kyoto 606-8502, Japan
}

\date{\today}

\begin{abstract}
The proposed space gravitational wave (GW) detector LISA has potential to detect stellar-mass black hole binaries (BBHs). The majority of the detected BBHs are expected to emit nearly monochromatic GWs, whose frequency evolution will  be efficiently described by Taylor expansions.  We study the measurability of the associated time derivative coefficients of the frequencies, by extending a recent work based on a simplified Fisher matrix analysis.  Additionally, we provide qualitative discussions on how to extract astrophysical information, such as orbital eccentricity and tertiary perturbation, from the observed derivative coefficients.
%nearly monochromatic 

%consolidate information to a finite number of parameters

%how they are related to binary parameters
%{known source direction}
 
\end{abstract}

\maketitle

\section{introduction}

The LIGO-Virgo-Kagra network had detected $O(100)$ binary black  holes (BBHs) by the end of   its O3 run \cite{KAGRA:2021vkt}. The proposed space GW detectors such as LISA \cite{LISA:2017pwj}, Taiji \cite{Ruan:2018tsw} and TianQin \cite{TianQin:2015yph} can observe similar   BBHs in the lower frequency regime \cite{2022arXiv220306016A,Sesana:2016ljz}.  

According to a recent statistical analysis employing the LVK GWTC-3 catalog,   LISA is expected to detect  $N_{\rm tot}\sim 2.3(T/{\rm 4yr})^{3/2} (\rho_{\rm thr}/10)^{-3}$ BBHs similar to those cataloged  \cite{Seto:2022xmh}  (see also \cite{Gerosa:2019dbe,Klein:2022rbf} for multiband observations).  Here $\rho_{\rm thr}$ is the detection threshold for the signal-to-noise ratio, and $T$ is the observational period.  The estimated number $N_{\rm tot}$ contains an uncertainty  of a factor   of two,  partly due to the unclear redshift dependence of the merger rate. 
  In the statistical  analysis,  the BBHs are simply  assumed to be nearly circular still  in the LISA band. 
  
In the total number $N_{\rm tot}$, the fraction of BBHs that will merge in the observational period $T=4$yr was estimated to be $\sim 5\%$  ($\sim 10\%$ for $T=10$yr) \cite{Seto:2022xmh}.  In fact, the majority of the detected BBHs  will be nearly monochromatic GW sources, with the peak number density (per logarithmic frequency interval) around  $f\sim5$mHz. 
Here the characteristic  frequency $\sim5$mHz is given as the tangential point of the noise  spectrum $S_n(f)$ with the slope $f^{-2/9}$ which is derived from the radiation reaction. In  \cite{Seto:2022xmh},  the median chirp mass of the detected  BBHs was also estimated to be $\mch\sim 30\so$ with a relatively large uncertainty for the distribution function at $\mch\gsim 50\so$. 

For the nearly monochromatic  GWs from these BBHs,
we will be able to efficiently describe  the frequency evolution,  using the time derivative coefficients   ${\dot f},{\ddot f},\cdots$ (also denoting $f^{(n)}=\p_t^n f$), as conventionally applied in similar occasions (e.g. spin-down of a pulsar).  This approach condenses the evolutionary information to a finite number of parameters and will be mathematically more  tractable, rather than directly dealing with the functional degrees of freedom. 
Meanwhile, 
stimulated by the recent reports on the long-term orbital evolution of    HM Cancri (an interacting double white dwarf binary at $f=6.2$mHz) \cite{2021ApJ...912L...8S,2023MNRAS.518.5123M}, the author  developed a simplified  Fisher matrix analysis on the estimation errors of the frequency derivatives up to the second order  ${\ddot f}$ \cite{Seto:2023skl} (see also \cite{Ebadi:2024oaq}).   In this paper, extending the analytical approach to include higher derivatives such as ${\dddot f}$ and beyond, we apply it to the nearly monochromatic BBHs in the LISA band.

Importantly, compared with the BBHs in the LVK band,  those in the LISA band could have much larger eccentricities \cite{Peters:1964zz,Seto:2016wom,Nishizawa:2016eza}.  Likewise, in the LISA band,  the environmental effects (e.g.  tertiary perturbation)  could be more strongly imprinted in the emitted GWs, because of the longer gravitational radiation timescale \cite{Seto:2008di,Seto:2016zfy,Tamanini:2018cqb,Wong:2019hsq,Kang:2021bmp,Robson:2018svj,Tamanini:2019usx,2021MNRAS.502.4199X,Zhang:2021pwe,Maeda:2023tao}.   
%It would be thus interesting to examine these signatures in the 
 In this paper, we also discuss the  possibilities of examining {the orbital eccentricities and the potential tertiary perturbations}  from the observables $f,{\dot f},{\ddot f},{\dddot f},\cdots$ (see also \cite{Takahashi:2023flk}).  To this end,  we introduce the non-dimensional parameters $C_n\propto f^{(n)} f^{n-1}/({\dot f})^n$ ($n=2,3,\cdots$) which are adjusted to  be $C_n=1$ for  an isolated circular BBH. We  apply  Fisher matrix results to evaluate the measurement errors for the parameters  $C_n$. 

To provide a concrete picture for our study, we  consider  an equal mass BBH  with the  characteristic values mentioned earlier: a chirp mass  $\mch=30\so$, a GW frequency $f=$5mHz  and a signal strength  $\rho=10$ (for $T=4$yr with LISA).  We set this BBH as our fiducial observational target. 
%We allow a mild eccentricity $e\lsim 0.1$ at $f=5$mHz.  

This paper is organized as follows. In section II, we evaluate the frequency derivatives $f^{(n)}$ for circular and eccentric BBHs.  We introduce  the non-dimensional measures $C_n$ and qualitatively argue the corrections due to  tertiary  perturbations.   In section III, based on the Fisher matrix  analysis, we present analytical expressions  for the estimation  errors of the time derivatives $f^{(n)}$.
In section IV, we discuss the measurability of these  derivative coefficients. We relatedly  discuss the prospects for studying  orbital eccentricity and  tertiary effects by using the quantities $C_n$. 
In  section V, we mention potential extensions of this study. Section VI is a brief summary of this paper. 

\section{evolution of  binary black holes}

\subsection{Circular Systems}

Let us first discuss the orbital evolution of a circular BBH. In the quadrupole approximation, the chirp rate of the GW frequency  $f$ (twice the orbital frequency) is given by  
\begin{equation}
\frac{df}{dt}=\frac{96  \pi^{8/3} G^{5/3} \mch^{5/3}  f^{11/3}}{5c^5  }\label{dfdt}
\end{equation}
with the chirp mass $\mch$, the gravitational constant $G$ and the speed of light $c$ \cite{Peters:1964zz}.  For  our  fiducial model parameters {(summarized in Table 1)} we have 
\begin{eqnarray}
{\dot f}=6.2 \times 10^{-13} \lmk \frac{\mch}{30\so} \rmk^{5/3}\lmk \frac{f}{\rm 5mHz} \rmk^{11/3} {\rm Hz\, s^{-1}} \label{df12}.
\end{eqnarray}
Correspondingly, in the observational period $T$,  the GW frequency $f$   increases only by 
$1.5(T/{\rm 4yr})$ percent.

\begingroup
\tabcolsep = 15.0pt
\def\arraystretch{1.3}
\begin{table}[]
\caption{Fiducial binary parameters and the associated  frequency derivatives due to the radiation reaction for a circular orbit. }
\begin{tabular}{|c|c|}
\hline
frequency $f$ & 5 mHz \\ \hline
chirp mass  $\mch$& 30 $\so$ \\ \hline
SNR $\rho$ (4yr)   & 10   \\ \hline\hline
$\dot f$ & $6.2\times 10^{-13} \rm Hz~s^{-1}$   \\ \hline
$\ddot f$ & $2.8\times 10^{-22} \rm Hz~s^{-2}$   \\ \hline
$\dddot f$ & $2.2\times 10^{-31} \rm Hz~s^{-3}$   \\ \hline
$\dddot f$ & $2.4\times 10^{-40} \rm Hz~s^{-4}$   \\ \hline
\end{tabular}
\end{table}
\endgroup

Recursively taking the time derivatives of Eq. (\ref{dfdt}), we can readily obtain the higher time  derivatives such as  
\begin{eqnarray}
{\ddot f}=2.8 \times 10^{-22} \lmk \frac{\mch}{30\so} \rmk^{10/3}\lmk \frac{f}{\rm 5mHz} \rmk^{19/3} {\rm Hz\, s^{-2}},
\end{eqnarray}
\begin{eqnarray}
{\dddot f}=2.2 \times 10^{-31} \lmk \frac{\mch}{30\so} \rmk^{5}\lmk \frac{f}{\rm 5mHz} \rmk^{9} {\rm Hz\, s^{-3}},
\end{eqnarray}
\begin{eqnarray}
{\ddddot f}=2.4 \times 10^{-40} \lmk \frac{\mch}{30\so} \rmk^{20/3}\lmk \frac{f}{\rm 5mHz} \rmk^{35/3} {\rm Hz\, s^{-4}}. \label{df42}
\end{eqnarray}
By integrating Eq.  (\ref{dfdt}), we can also estimate the time before the merger as 
\begin{eqnarray}
t_c=97 \lmk \frac{\mch}{30\so} \rmk^{-5/3}\lmk \frac{f}{\rm 5mHz} \rmk^{-8/3} {\rm yr}.
\end{eqnarray}
In terms of $t_c$, the time derivatives are roughly given by 
\begin{equation}
\p^n_t f=f^{(n)}\sim \frac{f}{t_c^n}.\label{scc1}
\end{equation}
More precisely,  from Eq. (1),  we can show 
\begin{eqnarray}
C_2\equiv \frac{3 {\ddot f} f}{11{\dot f}^2}&=&1,\label{c20}\\
C_3\equiv\frac{9 {\dddot f} f^2}{209{\dot f}^3}&=&1,\\
C_4\equiv\frac{ {\ddddot f} f^3}{209{\dot f}^4}&=&1.
\end{eqnarray}

Up  to now, we ignored the post-Newtonian effects.  The leading order  (1PN) correction to the product $C_2$ takes a negative value \cite{Blanchet:2013haa,Cutler:1994ys}
\begin{eqnarray}
-\frac2{11}\lmk \frac{743}{336}+\frac{11\eta}4 \rmk \lmk\frac{GM_t \pi f}{c^3} \label{pn1} \rmk^{2/3}
\end{eqnarray}
with the symmetric mass  ratio $\eta$ and  the total mass $M_t=\eta^{-3/5}\mch$.
  In  the LISA era,  from the preceding ground based detectors including the Einstein Telescope \cite{Punturo:2010zz}, we will be able to use the detailed prior information on the ratio $\eta$ for stellar mass BBHs and better model the PN corrections.  
For an equal mass binary (with  $\eta=1/4$), Eq. (\ref{pn1}) is expressed as 
\begin{eqnarray}
  -1.6 \times  10^{-4} \lmk  \frac{\mch}{30 \so} \rmk^{2/3}   \lmk  \frac{f}{\rm 5mHz} \rmk^{2/3} \label{pn2}.
\end{eqnarray}

\subsection{Eccentric Systems}

Next we discuss the time evolution of an eccentric  BBH (see also \cite{Tanay:2016zog}).   We continue to use $f$ as twice the orbital frequency.  At the quadrupole order, using the orbital eccentricity $e$,  we can write  down the evolution equations \cite{Peters:1964zz}
\begin{eqnarray}
\frac{df}{dt}&=&\frac{96  \pi^{8/3} G^{5/3} \mch^{5/3}  f^{11/3}}{5c^5   \left(1-e^2\right)^{7/2}}{ \left(1+\frac{73
   e^2}{24}+\frac{37 e^4}{96}\right)}\label{df}\\
   &\equiv&  F_1(f,e),
\end{eqnarray}
\add{
\begin{eqnarray}
\frac{de}{dt}&=&-\frac{304 \pi^{8/3} G^{5/3}f^{8/3}\mch^{5/3} e}{15c^5 (1-e^2)^{5/2}} \lmk  1+\frac{121e^2}{304}\rmk\label{de}\\
&\equiv&  E_1(f,e).
\end{eqnarray}
}
Applying the chain rule for derivatives, we can evaluate the second derivative $\ddot f$ as follows
\begin{eqnarray}
{\ddot f}=\frac{\p F_1}{\p f} F_1+\frac{\p F_1}{\p e} E_1\equiv F_2(f,e) \label{dd1}.
\end{eqnarray}

   The product $C_2$ in  Eq. (\ref{c20}) is now given as 
   \add{
\begin{eqnarray}
C_2&=&\frac{3 {\ddot f} f}{11{\dot f}^2}\label{c21}\\
%\end{eqnarray}
%We can derive its analytical expression as 
%\begin{eqnarray}
&= &\frac{4 \left(407 e^8+2344 e^6+93810 e^4+58720
   e^2+25344\right)}{11 \left(37 e^4+292
   e^2+96\right)^2}\nonumber\\
   & \equiv & {\CC}_2(e)\label{c21p}
\end{eqnarray}
}
with a perturbative expression 
\add{
\begin{eqnarray}
{\CC}_2(e)=1-\frac{2983e^2}{792}+O(e^4)\label{c2p}.
\end{eqnarray}
}
In contrast to Eqs.  (\ref{df}) and (\ref{de}),   the function ${\CC}_2(e)$  is regular  at $e=1$ with \add{
${\CC}_2(1)=4/11$}.

In  the same manner as Eq. (\ref{dd1}), we can evaluate the third derivative   $\dddot  f$ for an eccentric orbit as
\begin{eqnarray}
{\dddot f}=\frac{\p F_2}{\p f} F_1+\frac{\p F_2}{\p e} E_1\equiv F_3(f,e) \label{dd2}.
\end{eqnarray}
We then
  evaluate the product $C_3$   and define the function $\CC_3(e)$
  \add{
\begin{eqnarray}
{C}_3&=&\frac{9 {\dddot f} f^2}{209{\dot f}^3}\label{c32}\\
%\end{eqnarray}
%and derive 
%\begin{eqnarray} 
&= &\frac4{209 \left(37 e^4+292
   e^2+96\right)^3} (-9361 e^{12}+500532 e^{10}\nonumber\\
   & &-42206966
   e^8+145890160 e^6+129278336 e^4\nonumber\\
   & & +104147968
   e^2+46227456)\nonumber\\
   &\equiv & {\CC}_3(e)\label{c32p}
\end{eqnarray}}
with
\add{${\CC}_3(0)=1$,  ${\CC}_3(1)=20/209$ and 
\begin{eqnarray}
\CC_3(e)=1-\frac{2041e^2}{297}+O(e^4). \label{pt11}
\end{eqnarray}
}
Likewise, we can  evaluate the fourth derivative $\ddddot f(\equiv f^{(4)})$  and the product
 \begin{eqnarray}
C_4=\frac{ {\ddddot f} f^3}{209{\dot f}^4}. \label{c41}
\end{eqnarray}
The analytic expression for the corresponding function $\CC_4(e)$ is given as \add{
\begin{eqnarray}
\CC_4(e)&=&\frac{8}{5643 (37 e^4+292
   e^2+96)^4} (421245 e^{16}\nonumber\\
   & & -100387440 e^{14}+12951727630
   e^{12}\nonumber\\
   & & -122186028060 e^{10}+117746520816
   e^8\nonumber\\
   & & +132600265216 e^6+150472180736 e^4\nonumber\\
   & & +137985376256
   e^2+59910782976) \label{c42p}
\end{eqnarray}
with $\CC_4(0)=1$, $\CC_4(1)=40/1881$ and 
\begin{eqnarray}
K_4(e)=1-\frac{1898287e^2}{192456}+O(e^4).
\end{eqnarray}
}
In Fig. 1, we present the functions $\CC_2(e)$, $\CC_3(e)$ and $\CC_4(e)$.  All of them \add{decrease}  monotonically with $e$. By evaluating the products $C_n$ from the observed frequency derivatives, we might inversely estimate the orbital eccentricity $e$.

\begin{figure}
 \includegraphics[width=.95\linewidth]{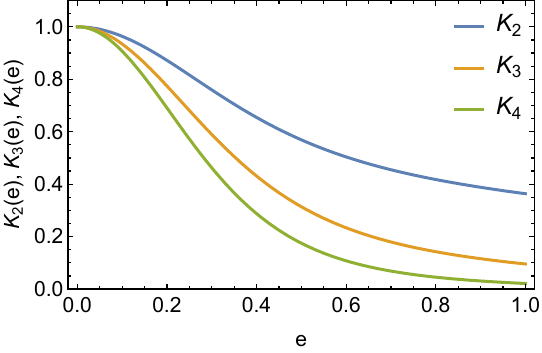} 
 \caption{The three functions ${\CC}_2(e)$, $\CC_3(e)$ and $\CC_4(e)$ defined in Eqs. (\ref{c21p}) (\ref{c32p}) and (\ref{c42p}).  For  a circular binary ($e=0$),  we have $\CC_2=\CC_3=\CC_4=1$. At $e=0.1$, we have \add{$\CC_2=0.96$,  $\CC_3=0.93$ and $\CC_4=0.91$}.
 }  \label{fig:volume}
\end{figure}

The PN corrections  to the products $C_n$ depend  on the eccentricity $e$.  In this paper, we approximately use the results  (\ref{pn2}) also for mildly eccentric BBHs.

\subsection{Tertiary Effects}
{Here, we briefly discuss  the GW phase modulation induced by  a tertiary (of mass $m_s$) around a BBH (see, e.g.  \cite{Seto:2023rfh,Bassa:2016fiy,Kaplan:2016ymq} for the effects on the derivative coefficients $f^{(n)}$), as a representative environmental perturbation.
In this subsection, we temporarily  attach the subscript $s$ to the tertiary related quantities and $r$ to the radiational effects (under the approximation ignoring the cross terms \cite{Seto:2023rfh}). 
  For simplicity, we assume that the outer orbits is hierarchical and circular with the period $P_s$ and the semimajor axis $R_s$.
  We have 
   \begin{equation}
R_s=\lkk  \frac{G(M_t+m_s)P_s^2}{4\pi^2}   \rkk^{1/3}.
\end{equation} 
from the third Kepler law. 
   In this simplified case, the BBH moves approximately  on  a circular orbit around the barycenter of the triple system. 
  } 
   
  { Following \cite{Seto:2023rfh}, we briefly discuss the tertiary corrections  (equivalent to the R{\o}mer effects) to the apparent GW frequency and its derivative coefficients. The positional variation induces the phase shift as    
  \begin{equation}
\Phi_s(t)=-2\pi f_r D(t)/c  \label{ps1}
\end{equation}
  where $D(t)$ is the line-of-sight distance variation and expressed as ($\phi_s$: the initial outer  orbital phase)
     \begin{equation}
D(t)=D_s \cos [2\pi  t/P_s+\phi_s]
\end{equation}
  with the projected distance
    \begin{equation}
D_s=\frac{m_s}{M_t+m_s} R_s\sin I_s
\end{equation} 
  ($I_s$: the outer inclination angle). 
  The tertiary corrections to the apparent GW frequency and its time derivatives are given as
    \begin{equation}
f_s=\frac1{2\pi}\frac{d\Phi_s}{dt},~~{\dot f}_s=\frac1{2\pi}\frac{d^2\Phi_s}{dt^2},~~{\ddot f}_s=\frac1{2\pi}\frac{d^3\Phi_s}{dt^3} \label{ps3}
\end{equation} 
 and so on. }
  
  We put $v_s\sim 2\pi D_s/P_s$ as the line-of-sight component of the outer  circular velocity of the binary.  Then, for an observational time $T(\ll P_s)$,  the tertiary effects  $f_s^{(n)}$ have the  characteristic magnitudes  
\begin{eqnarray}
|f_s^{(n)}|\sim f  \lmk \frac{v_s}c\rmk   \lmk  \frac{2\pi}{ P_s} \rmk^n. \label{scc12}
\end{eqnarray}
%Here  we dropped the factors related to the outer orbital phase (see e.g. XX for $P_s\lsim T$). 

A typical field system will have $v_s/c\ll 1$.  Depending on the outer orbital phase, the tertiary corrections $f_s^{(n)}$ can take both  signs.    

%{need revision, smooth the flow  (subscript etc)}
Now we compare the tertiary and radiational effects, using Eqs. (\ref{scc1}) and (\ref{scc12}).  We can roughly evaluate the ratios 
\begin{eqnarray}
\frac{|f_s^{(n)}|}{f_r^{(n)}}\sim   \lmk \frac{v_s}c\rmk   \lmk  \frac{2\pi t_{c}}{ P_s} \rmk^n   \label{r24}
\end{eqnarray}
with the time $t_c$  before the merger.  
For  $2\pi t_c/P_s\gg 1$, 
the tertiary effects could be relatively more important  at larger $n$.  However, in  contrast to Galactic white dwarf binaries   ($t_c\gsim 10^{5}$yr  in the LISA band), our BBHs have much shorter merger time  $t_c\sim 100$yr, and the amplification  factor $ \lmk  {2\pi t_{c}}/{ P_s} \rmk$  will work less efficiently in Eq. (\ref{r24}).

As a concrete reference model, we  set the tertiary mass $M_s\sim 25\so$  and an edge-on outer orbit ($I_s=\pi/2$) of $P_s=60$yr ($\ll 2\pi t_c$). 
%(not comfortably satisfying the condition $T\ll P_s$), 
Then, using Eqs. (\ref{ps1})-(\ref{ps3}),  we obtain 
\begin{eqnarray}
\frac{{f}_s}{{ f}_r}&=&1.1\times 10^{-4}\sin\psi_s ,\label{rr0}\\
\frac{{\dot f}_s}{{\dot f}_r}&=&8.2\times 10^{-4}\cos\psi_s ,\label{rr1}\\  \frac{{\ddot f}_s}{{\ddot f}_r}&=&-6.0\times 10^{-3}\sin\psi_s,  \label{rr2} \\
 \frac{  {\dddot f}_s}{{\dddot f}_r}&=&-2.6\times 10^{-2}\cos\psi_s  \label{rr3}
\end{eqnarray}
%$|{\dot f}_s|/{\dot f}_r=8.2\times 10^{-4}$,  $|{\ddot f}_s|/{\ddot f}_r=6.0\times 10^{-3}$ and   $|{\dddot f}_s|/{\dddot f}_r=2.6\times 10^{-2}$
with the outer orbital phase parameter $\psi_s=2\pi t/P_s+\phi_s$.
 %(also taking into account the relevant numerical factors on the right-hand-side of 
 %Eq. (\ref{r24})) \cite{Seto:2023rfh}.  

In this paper, we ignore the dynamical interaction between the inner and outer orbits (see, e.g. \cite{Samsing:2024syt}).  This would be a reasonable approximation for our reference tertiary system, at least for the Kozai-Lidov mechanism. 

If the tertiary effects are much smaller than the radiational effects (as for the reference model above), we can 
perturbatively deal with the former. {For $n=2$, we have
\begin{equation}
C_2=\frac{3 ({\ddot f}_r+{\ddot f}_s) (f_r+f_s)}{11({\dot f}_r+{\dot f}_s)^2}\simeq K_2(e) \lmk 1+\frac{ f_s}{ f_r}-2 \frac{\dot f_s}{\dot f_r}   +\frac{\ddot f_s}{\ddot f_r}  \rmk
\end{equation}
 and similarly 
\begin{equation}
C_n\simeq \CC_n(e) \lmk 1+(n-1)\frac{ f_s}{ f_r}-n \frac{\dot f_s}{\dot f_r}   +\frac{f_s^{(n)}}{f_r^{(n)}}    \rmk  \label{cter}
\end{equation}
with the inequalities  \add{$\CC_n(e)\le K_n(0)= 1$}  from the definitions of $\CC_n(e)$.  As shown in Eq. (\ref{cter})
The eccentricity and tertiary effects degenerate in the product $C_n$. 
}

For a  nearly circular binary $e\sim 0$, we could have a chance to observe  \add{the excess $C_n-1>0$} due to the second  factor in Eq. (\ref{cter}), depending on the outer orbital phase $\psi_s$.   {This inequality cannot be realized for an isolated eccentric BBH and suggests the existence  of a third body, as one of its causes. }
% Conversely,  if we observationally get $C_n< 1$,  the  tertiary perturber could be an interesting candidate for the cause.  

The situation will be more complicated for an eccentric binary, for which the tertiary effect might be actually more interesting in view of dynamical processes. 
In some cases,  we will be able to  measure the higher order product $C_3$ in addition to the lowest one $C_2$. Then the orbital eccentricity $e$ can be estimated independently for $n=2$ and 3 by $e=K^{-1}_n(C_n)$.  The tertiary  effects could induce a mismatch  between the inverted values. 
%While the small tertiary effects could  be potentially interesting, we put aside them mostly  in  our analysis below.  
Note that we have $K'_n(0)=0$, and the inversions work less efficiently for nearly circular binary $e\simeq0$.

Below, in particular for the expected values for the frequency derivatives $f^{(n)}$, we mainly deal with the radiational effects, dropping the subscripts $r$ and $s$ (if unnecessarily). We will return to the tertiary effect later in Sec. IV D.

\section{fisher matrix  analysis}

In this section, we apply the Fisher matrix  analysis to analytically evaluate how well we can measure the  frequency derivatives  in a matched filtering analysis for a nearly monochromatic BBH.

\subsection{Simplified Waveform Model}
We basically extend the earlier work by the author \cite{Seto:2023skl} (see also \cite{Seto:2002dz,Robson:2018svj}) and apply the simplified waveform model 
\begin{eqnarray}
h(t)=A\cos[\Phi_n(t)] \label{ph}
\end{eqnarray}
with the constant amplitude $A$ and the phase function
\begin{eqnarray}
\Phi_n(t)=2\pi \lmk ft+\frac{{\dot f}t^2}{2!}+\frac{{\ddot f}t^3}{3!} %+\frac{{\dddot f}t^4}{4!} 
+\cdots +\frac{ f^{(n)}t^{n+1}}{(n+1)!}   \rmk %\nonumber\\
+\varphi. \nonumber\\ \label{cub}
\end{eqnarray}
This phase function is  
perturbatively expanded at the time origin $t=0$.  Here the index $n$ represents the order of the highest  frequency derivative $f^{(n)}$ (different from  the associated power index of  $t^{n+1}$).

In this paper, 
 we take the time origin $t=0$ (for the expansion (\ref{cub})) at the midpoint of the observational  period  $T$.  {More specifically, we set our observational period by $t\in[-T/2,T/2]$ not by $t\in [0,T]$.  As we see below, this choice is for reducing some of the Fisher matrix elements by a block diagonalization.}  Applying the Parseval's  theorem for a nearly monochromatic signal, we obtain the expected signal-to-noise ratios $\rho$ as
\begin{eqnarray}
\rho^2&=&\frac2{S_{\rm n}(f)}\int_{-T/2}^{T/2}  h(t)h(t)dt \label{inner}\\
&=&\frac{A^2 T}{S_{\rm n}(f)}  \label{snr}
\end{eqnarray}
with the effective noise spectrum  $S_n(f)$.
We have ignored the time dependence  of the confusion noise, which will be  unimportant at $f\gsim 5$mHz and  $T\gsim4$yr \cite{Robson:2018ifk}.

Next, we apply  the Fisher matrix  analysis to evaluate the measurement errors for the fitting  parameters ${\vect \theta}_n=\{\varphi,f,{\dot f},\cdots, f^{(n)}    \} $ in the   phase function (\ref{cub}).  Note that the error for the  amplitude $A$ has no correlation with those for  the parameters ${\vect \theta}_n$.  The Fisher matrix  elements are formally  given as
\begin{eqnarray}
F_{\theta_i\theta_j}=\frac2{S_{\rm n}(f)}\int_{-T/2}^{T/2} \p_{\theta_i }h(t) \p_{\theta_j}h(t)  dt.\label{fis}
\end{eqnarray}
We can analytically evaluate  these matrix  elements and replace the common factor $A^2/S_n(f)$ by $\rho^2/T$.  The covariance matrix  $\lla \delta{\theta_i} \delta{\theta_j}  \rra$  for  the measurement errors $\delta \theta_i$ can be estimated  by the  inverse of the Fisher matrix as 
\begin{eqnarray}
\lla \delta{\theta_i} \delta{\theta_j}  \rra=(F^{-1})_{\theta_i\theta_j}. \label{inv1}
\end{eqnarray}
Below, for notational simplicity, we put the rms  value of the measurement error by 
\begin{eqnarray}
\Delta{\theta_i}\equiv \lla \delta{\theta_i} \delta{\theta_i}  \rra^{1/2}.\label{inv2}
\end{eqnarray}

Here, importantly,  the Fisher matrix and its inverse matrix  have  block diagonal structures.
We can divide the parameters ${\vect \theta}_n$ into two subsets, on the basis of the evenness and oddness  of the time derivatives  (effectively regarding $\varphi$ as  the minus first derivative).   The odd ones  are $\{\varphi,{\dot f},{\dddot  f},\cdots  \}$ and the even ones are  $\{f,{\ddot f},\cdots  \}$. The two groups have no error correlation,  owing to  the symmetric cancellations in the time integrals (\ref{fis}).  For example, we have
\begin{eqnarray}
F_{\varphi  {\ddot f}}\propto \int_{-T/2}^{T/2} t^3 dt=0.\label{cancel}
\end{eqnarray}
Here our midpoint expansion plays the critical role. 

Our simplified  waveform model (\ref{ph}) does not contain other  parameters such as the source direction and orientation. In particular,  the phase function (\ref{cub}) lacks the Doppler phase modulation induced by the annual rotation  of the detector around the Sun.    Therefore, one might  be skeptical to the validity of our approach. These concerns were closely examined in \cite{Seto:2023skl} by comparison of our analytical  approach  with the full Fisher matrix analysis  including other fitting parameters and the Doppler and amplitude modulations \cite{Cutler:1997ta,Takahashi:2002ky}. It was shown that,  for observational period $T\gsim 2$yr,  our simplified evaluations work quite efficiently for  the phase related parameters ${\vect \theta}_2=\{\varphi,f,{\dot f},{\ddot f}    \} $.  This is because,    at $T\gsim 2$yr, we can distinguish the periodic Doppler  modulation  from  the perturbative frequency drift (\ref{cub})  truncated at a finite order.  As a result, the mutual correlation between the source geometric parameters and those in Eq. (\ref{cub}) becomes weak, and our simple approach will be valid even for $n=4$. 
Note that,  For $T\gsim2$yr, the cancellations similar to Eq. (\ref{cancel}) effectively work, even if we include the annual amplitude modulation  \cite{Seto:2023skl}.

\subsection{Analytical Results}
In this subsection,  for various orders $n$ of  the phase expansion, we present the concrete expressions  for the estimation errors of  the fitting parameters  ${\vect \theta}_n=\{\varphi,f,{\dot f},\cdots, f^{(n)}    \} $.   As mentioned in the previous subsection,   these expressions will be valid   for $T\gsim  2$yr. 

If  the apparent frequency evolution is well described by the second-order expansion  $\Phi_2(t)$ with the four parameters  ${\vect \theta}_2=\{\varphi,f,{\dot f},{\ddot f}\} $,  we can obtain the Fisher matrix results 
\begin{eqnarray}
\Delta\varphi=\frac3{2\rho},~\Delta f=\frac{5\sqrt3}{2\pi \rho T}, ~\Delta {\dot f}=\frac{6\sqrt5}{\pi \rho T^2},
~\Delta {\ddot   f}=\frac{60\sqrt7}{\pi \rho T^3} \label{v2},
\end{eqnarray}
as presented in  \cite{Seto:2023skl}.
For the third-order expansion $\Phi_3(t)$ with the five fitting   parameters  ${\vect \theta}_3=\{\varphi,f,{\dot f},{\ddot f},{\dddot f}\}$ (newly including $\dddot f$), we obtain
\begin{eqnarray}
\Delta\varphi&=&\frac{15}{8\rho},~\Delta f=\frac{5\sqrt3}{2\pi \rho T}, ~\Delta {\dot f}=\frac{21\sqrt5}{\pi \rho T^2},~\Delta {\ddot   f}=\frac{60\sqrt7}{\pi \rho T^3} \nonumber  \\
\Delta {\dddot   f}&=&\frac{2520}{\pi \rho T^4}. \label{v3}
\end{eqnarray}
Because of the block diagonalization,  the results $\{ \Delta f, \Delta{\ddot f}\}$ are the same as those in Eq. (\ref{v2}). In contrast, the remaining  ones $\{\Delta \varphi,\Delta {\dot  f} \}$ become larger, reflecting their correlation with the additional error $\Delta {\dddot f}$.

Next we  move to  the fourth-order  function  $\Phi_4(t)$ with the six  fitting parameters ${\vect \theta}_4=\{\varphi, f,{\dot f},{\ddot f},{\dddot f},{\ddddot f}\}$.  We obtain
\begin{eqnarray}
\Delta\varphi&=&\frac{15}{8\rho},~\Delta f=\frac{35\sqrt3}{8\pi \rho T}, ~\Delta {\dot f}=\frac{21\sqrt5}{\pi \rho T^2},~\Delta {\ddot   f}=\frac{270\sqrt7}{\pi \rho T^3} \nonumber  \\
\Delta {\dddot   f}&=&\frac{2520}{\pi \rho T^4},
~\Delta {\ddddot   f}=\frac{15120 \sqrt{11}}{\pi \rho T^5}.  \label{v4}
\end{eqnarray}
Here the results  $\{\Delta \varphi,\Delta {\dot  f}, \Delta {\dddot f} \}$ are the same as those in  Eq.(\ref{v3}), but others $\{\Delta  f, \Delta {\ddot  f}\}$ become larger.   In this manner, under the midpoint expansion, we can append the highest derivative coefficients $f^{(n+1)}$ to the fitting parameters, without sacrificing the accuracy of the second highest one  $f^{(n)}$.  As we see in the next section, this will be preferable at actual data analysis for  nearly monochromatic BBHs. 

{
In Fig. 2, to show the dependence on the order of the phase expansion ($n=2,3$ and 4), we compare the numerical coefficients for the errors presented in Eqs. (\ref{v2})-(\ref{v4}) normalized by  those presented in Eq. (\ref{v4}). We can see that the lower order parameters such as $ \varphi$ are less affected by the order of the derivative expansion.  This seems reasonable, given the differences of the power indexes in the phase function (\ref{cub}).  
\begin{figure}
 \includegraphics[width=.95\linewidth]{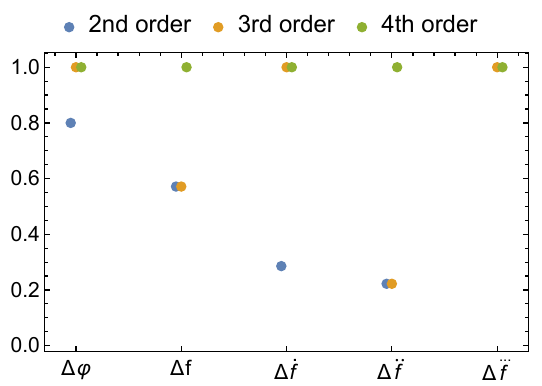} 
 \caption{Comparison of the numerical factors in  Eqs. (\ref{v2})-(\ref{v4})  normalized by those  in Eq. (\ref{v4}).
 }  \label{fig:a1}
\end{figure}
}

So far, we have applied the GW phase expansion (\ref{cub}) at the midpoint $t=0$ of the observation period $[-T/2,T/2]$.  
For comparison, we evaluated the Fisher matrix results,  expanding the GW phase   at the initial point $t=0$ of the integration period $[0,T]$.  For the six  fitting parameters ${\vect \theta}_4=\{\varphi, f,{\dot f},{\ddot f},{\dddot f},{\ddddot f}\}$, we obtained
\begin{eqnarray}
\Delta\varphi&=&\frac{6}{\rho},~\Delta f=\frac{35\sqrt3}{\pi \rho T}, ~\Delta {\dot f}=\frac{336\sqrt5}{\pi \rho T^2},~\Delta {\ddot   f}=\frac{2160\sqrt7}{\pi \rho T^3} \nonumber  \\
\Delta {\dddot   f}&=&\frac{25200}{\pi \rho T^4},
~\Delta {\ddddot   f}=\frac{15120 \sqrt{11}}{\pi \rho T^5}.  \label{v42}
\end{eqnarray}
The magnitude  $\Delta {\ddddot   f}$  for the highest derivative is identical  to Eq. (\ref{v4})  for the midpoint expansion (as explained below), but other parameters have much larger estimation errors.  {
Note that, in the case of the initial point expansion, the diagonal elements $F_{f^{(n)}f^{(n)}}$ of the Fisher  matrix (\ref{fis}) are $2^{2n+2}$ times larger than those for the midpoint expansion (again regarding $\varphi\propto f^{(-1)}$). However,  the diagonal elements of the inverted matrix (see Eqs. (\ref{inv1}) and (\ref{inv2})) become generally larger for the initial point expansion, as shown in Eqs. (\ref{v4})-(\ref{v42}). This is  due to the off-diagonal elements of the Fisher matrix (related to correlations of the parameters), and the block diagonal structure of the midpoint expansion would play an important role. }

{Eqs. (\ref{v4}) and (\ref{v42}) are obtained under the assumption that the phase function $\Phi(t)$ is well approximated by the fifth order polynomial expansion of the time variable $t$. Using the difference of the time origin $t=0$, 
 we can write down the third derivative coefficient  of the initial point expansion
\begin{equation}
{\dddot f}\simeq {\dddot f}_{\rm m}-\frac{T}2 {\ddddot f}_{\rm m } 
\end{equation}
in terms of these evaluated under the midpoint expansion ${\dddot f}_{\rm m }$ and ${\ddddot f}_{\rm m }$ (temporarily adding the subscript m). Then, we have
\begin{equation}
\Delta {\dddot f}\simeq \lmk \Delta {\dddot f}_{\rm m}^2+\frac{T^2}4 \Delta {\ddddot f}_{\rm m }^2\rmk^{1/2}
\end{equation}
due to the block diagonalization for the midpoint expansion. We can easily confirm the expression $\Delta {\dddot f}$ in (\ref{v42}) from Eq. (\ref{v4}).  We also have ${\ddddot f}={\ddddot f_{\rm m }}$ and thus $\Delta {\ddddot f}=\Delta {\ddddot f_{\rm m }}$, as commented earlier. 
} 

\subsection{Order  of the Expansion}

In any case, we should carefully adjust the order $n$ of the phase expansion at the matched filtering analysis.  An unnecessarily large $n$ results in avoidable degradation of the important parameters such as $\ddot  f$ (see Fig. 2).  At the same time,  they will be biased,  if we take a too small $n$.   
In reality, we should  adjust the order $n$, also taking into account the observed properties  of the actual BBH samples detected by space detectors.  Below, as a working hypothesis, we  stop the expansion at the lowest order $n$ satisfying the resolution condition
\begin{equation}
f^{(n)} \lsim  \Delta f^{(n)}.  \label{cnd}
\end{equation}
Here, as the reference values,  we use the derivatives $f^{(n)}$ given in Eqs. (\ref{df12})-(\ref{df42})  for radiation reaction.
At the next order expansion $n+1$,  for our fiducial BBHs,  we will have $f^{(n+1)} \ll  \Delta f^{(n+1)}$   from the scaling relations $f^{(n)}\propto t_c^{-n}$ and $ \Delta f^{(n)}\propto T^{-n+1/2}$, together with the increasingly large numerical  factors of the latter (see Eqs. (\ref{v2})-(\ref{v4})).  

%skip every other one

In Fig. 3, assuming circular BBHs and using expressions in Secs. II A and III B,   we present the integration time $T$ satisfying the condition $\Delta f^{(n)}=f^{(n)}$ for various chirp masses $\mch$ at $f=5$mHz.
For each index $n$, we use  the error $\Delta f^{(n)}$ obtained with  the corresponding expansion $\Phi_n(t)$ (e.g.  $\Delta {\ddot f}$  from  Eq.  (\ref{v2})   not from  Eq. (\ref{v4})). We can  roughly understand the required order $n$ of the expansion  as a function of the integration time $T$.

In this figure,  considering the applicability of our simplified phase model (\ref{ph}), the results will be valid at $T\gsim2$ yr. 
For  the Fisher matrix, we have also used the nearly monochromatic approximation for evaluating the time integrals (\ref{fis}).      Therefore, in Fig. 3, we  show the time $T$ during which  the GW frequency $f$ changes by  1\% and 5\% 
 (namely ${\dot f} T/f=0.01 $  and  $0.05 $). In Fig. 4, we preset a similar plot, now fixing the chirp mass at $\mch=30\so$.

\begin{figure}
 \includegraphics[width=.95\linewidth]{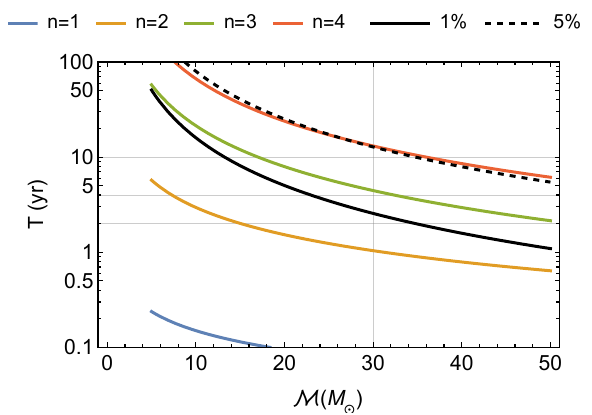} 
 \caption{The integration time $T$  satisfying the condition $\Delta f^{(n)}=f^{(n)}$  at $f=5$mHz. The results are given for circular BBHs which have the signal-to-noise ratio $\rho=10$ in  $T=4$yr.  The GW frequency $f$ changes by  1\% and 5\% during the time $T$  shown by the black lines.
 }  \label{fig:volume}
\end{figure}

\begin{figure}
 \includegraphics[width=.95\linewidth]{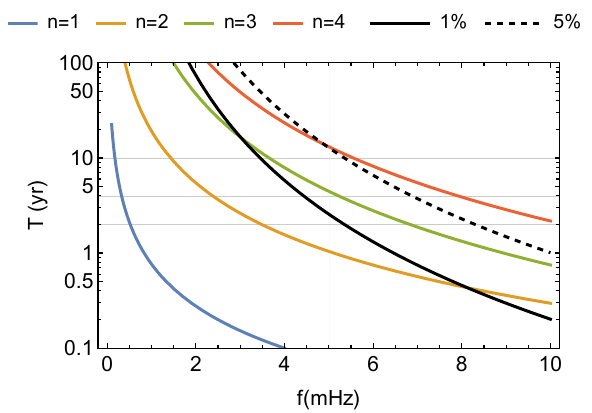} 
 \caption{The integration time $T$  satisfying the condition $\Delta f^{(n)}=f^{(n)}$   at various GW frequency $f$. The results are given for circular BBHs with the fixed chirp mass $\mch=30\so$  and  $\rho=10$ in  $T=4$yr.  %The GW frequency $f$ changes by  1\% and 5\% during the time $T$  shown by the black lines. 
 The results at $f=5$mHz are identical to those for $\mch=30\so$ in Fig. 3. 
 }  \label{fig:volume}
\end{figure}

\begin{figure}
 \includegraphics[width=.9\linewidth]{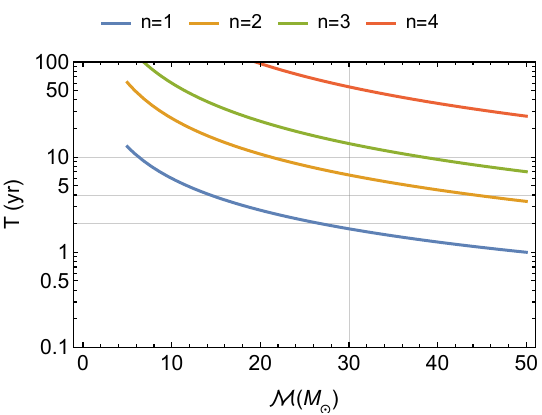} 
 \caption{The integration time $T$  satisfying the condition (\ref{pe}) for the phase mismatch.  The results are given for circular BBHs which have the GW frequency $f=5$mHz and various chirp masses $\mch$.  %The GW frequency $f$ changes by  1\% and 5\% during the time $T$  shown by the black lines. 
 }  \label{fig:volume}
\end{figure}

\begin{figure}
 \includegraphics[width=.9\linewidth]{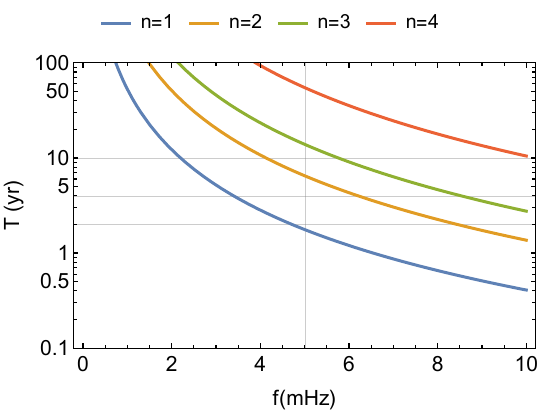} 
 \caption{The integration time $T$  satisfying the condition (\ref{pe}) for the phase mismatch.   The results are given for circular BBHs which have the fixed chirp mass $\mch=30\so$ various  GW frequencies. %The GW frequency $f$ changes by  1\% and 5\% during the time $T$  shown by the black lines. 
 The results at $f=5$mHz are identical to those for $\mch=30\so$ in Fig. 5. 
 }  \label{fig:volume}
\end{figure}

 {We have assumed to determine the order $n$ of the Taylor expansion, following the condition (\ref{cnd}).  This choice is equivalent to set $f^{(n+1)}=f^{(n+2)}=\cdots=0$ for the higher derivative coefficients.  The associated phase mismatch is roughly given by  
 \begin{eqnarray}
 \frac{f^{(n+1)}}{(n+2)!}t^{n+2}, \label{pe0}
 \end{eqnarray}
which takes the maximum magnitude at the initial and end points $t=\pm T/2$. From the viewpoint of matched filtering analysis, the phase mismatch due to the incomplete template should be much less than unity (see e.g. \cite{Creighton:2011zz}).}

{Here we evaluate the  integration time $T$ with which the phase  mismatch becomes unity as
 \begin{eqnarray}
 \frac{f^{(n+1)}}{(n+2)!}\lmk \frac{T}2  \rmk^{n+2}=1. \label{pe}
 \end{eqnarray}
 The results are presented in Figs. 5 and 6. Comparing Figs. 3 and 5, we can see that, for a given $n$ (e.g. $n=3$), the integration time $T$ in Fig. 3 is generally smaller than that in Fig. 5.  Therefore, following the condition (\ref{cnd}),  the associated phase mismatch will be much less than unity.  We can confirm similar results in Figs. 4 and 6.  We should also note that the left-hand side of Eq. (\ref{pe}) is proportional to $f^{(n+1)}T^{(n+2)}$, in a  similar way  to the ratio $f^{(n+1)}/(\Delta f^{(n+1)})$.
}

\section{observational prospects}

We now discuss the prospects for measuring the frequency derivatives for our fiducial BBH.   For a circular orbit, from Eqs. (\ref{df12})-(\ref{df42}) (also Table I),  we have 
\begin{eqnarray}
{\dot f}=6.2 \times 10^{-13}{\rm Hz\,s^{-1}}, \\
{\ddot f}=2.8 \times 10^{-22}{\rm Hz\, s^{-2}},\label{ddp}
\\{\dddot f}=2.2 \times 10^{-31}{\rm Hz\,s^{-3}},\label{dddp}
\\{\ddddot f}=2.4 \times 10^{-40}{\rm Hz\,s^{-4}}.
\end{eqnarray}
Below, also for mildly eccentric BBHs ($e\lsim 0.1$),   we use these results as the  reference values. 

In this section, we mainly set  $T=4$ yr and $\rho=10$, given the nominal operation period of LISA. In section IV D, we discuss the case for an extended case $T=10$ yr and $\rho=(10/4)^{1/2}\times 10\simeq 16$.

\subsection{Estimation Errors}
Using expressions (\ref{v2}) for the  second order phase expansion  $\Phi_2(t)$, we obtain 
\begin{eqnarray}
\Delta {\dot f}=2.7\times 10^{-17} {\rm Hz\,s^{-1}}\label{211}\\
\Delta {\ddot f}=2.5\times 10^{-24} {\rm Hz\,s^{-2}}\label{221}
\end{eqnarray}
for $\rho=10$ and $T=4$yr.  Given the relation 
$\Delta {\ddot f}\ll {\ddot  f}$ for the highest derivative $\ddot f$ (presented in Eq. (\ref{ddp})),  we examine  the next-order function $\Phi_3(t)$.  From expression (\ref{v3}), we obtain 
\begin{eqnarray}
\Delta {\dot f}=9.4\times 10^{-17} {\rm Hz\,s^{-1}}\label{2212}\\
\Delta {\ddot f}=2.5\times 10^{-24}{\rm Hz\,s^{-2}}\label{222}\\
\Delta {\dddot f}=3.2\times 10^{-31}{\rm Hz\,s^{-3}}\label{232}.
\end{eqnarray}
Now we have a relation 
$ {\dddot  f} \lsim \Delta {\dddot f}$ for the highest derivative.
As shown in Eqs. (\ref{221}) and (\ref{222}), due to the block diaginalization, the error $\Delta {\ddot f}$ is invariant. In contrast,  the error $\Delta {\dot f}$ in Eq.  (\ref{2212}) becomes 7/2 times  larger than that in Eq. (\ref{211}). We still have $\Delta {\dot f}\ll {\dot f}$ and this degradation of the factor 7/2 might not be a practial problem from astrophysical point of views.   This is because we usually do not need to very accurately (sub-percent level) measure the rate ${\dot f}$ (or the redshifted chirp mass) alone.
When we extract some astrophysical information by combining the rate  $\dot f$ with other fitting parameters (e.g. $\ddot f$ for the combination $C_2$ as discussed in section II B), the error $\Delta {\dot f}$ will have a minor contribution to the total error. 

Using Eq. (\ref{v4}) for the fourth-order expansion $\Phi_4(t)$, we have the estimation errors
\begin{eqnarray}
\Delta {\dot f}=9.4\times 10^{-17} {\rm Hz\,s^{-1}}\\
\Delta {\ddot f}=1.1\times 10^{-23}{\rm Hz\,s^{-2}}\label{233x}\\
\Delta {\dddot f}=3.2\times 10^{-31}{\rm Hz\,s^{-3}}\label{233}\\
\Delta {\ddddot f}=5.0\times 10^{-38}{\rm Hz\,s^{-4}}.
\end{eqnarray}
  We now have 
${\ddddot  f}\ll \Delta {\ddddot f}$ for the highest derivative,  and this expansion will be excessive in view of the criteria (\ref{cnd}), hampering  the resolution  $\Delta {\ddot f}$ (increased by a factor of 9/2 from Eq. (\ref{222})).  The reasonable choice will be  $n=3$, and we below  use Eqs. (\ref{2212})-(\ref{232}) for the estimation errors. 

\subsection{Eccentricity Estimation}

We next discuss the extraction of the astrophysical information from the observed products $C_n$. 

 Given the hierarchy of the relative errors $\Delta f^{(n)}/f^{(n)}$, the measurement  error $\Delta C_n$  is estimated   to be 
\begin{eqnarray}
 \Delta C_n\sim \frac{\Delta C_n}{C_n}\sim   \frac{\Delta {f^{(n)}}}{ f^{(n)}}
\end{eqnarray}
up to moderate eccentricity with $C_n\sim 1$. 
 For our fiducial BBH, from Eqs. (\ref{ddp}) and (\ref{222}), we have $\Delta C_2\sim 9\times 10^{-3}$.       With the relation 
 $\Delta {\dddot f}\sim {\dddot f}$ for the next order one $C_3$,  we cannot measure  $C_3$ in  real terms.  In addition, even for the reference system in Sec. II  C,  the shift of $C_2$ due to  the tertiary  is comparable to its error $\Delta C_2$.  Therefore, below  in this subsection, we concentrate on the eccentricity measurement from the product $C_2$. 

If the observed value $C_2$ is consistent  with unity (namely $|C_2-1|\lsim \Delta C_2$), we can apply Eq. (\ref{c2p}) and set an upper limit to the eccentricity as  
\begin{eqnarray}
e\lsim \lnk   \frac{\add{792}}{2983}\Delta C_2 \rnk^{1/2}.
\end{eqnarray}
We have 
$e\lsim0.0\add{5} $  for our fiducial BBH.

In contrast, if the product $C_2$ satisfies the condition $C_2-1\add{\gg} \Delta C_2$, we can solve Eq.  (\ref{c21p}) for the eccentricity $e$ with the estimation error 
\add{\begin{eqnarray}
 \Delta e\sim \frac{\Delta C_2}{|\p_e\CC_2(e)|}. \label{ere}
\end{eqnarray}
}

As an example, for the true value $e=0.1$, we have $C_2=\CC_2(0.1)=\add{0.96}$, and the expected estimation error
  becomes  \add{$\Delta e\sim1.3\times  10^{-2}$}.

%%%%%%%%%%%%%%%%%%%%%%%%%%%%%%%%
\subsection{Extension to 10yr}

In this subsection, we discuss the results  for an extended  observation of $T=10$ yr.  Compared with $T=4$ yr, the signal-to-noise  $\rho$ increases  by  a  factor of $2.5^{1/2}\sim1.6$ and the estimation errors  $\Delta f^{(n)}$ formally  shrink by $ 0.4^{n+3/2}$ (0.1 for $n=2$, 0.016 for $n=3$ and 0.006 for $n=4$).   However,  for a longer $T$, we need to increase the order $n$ of the phase expansion, following the criteria (\ref{cnd}).    Therefore,    these  scaling factors cannot be simply applicable.

\begingroup
\tabcolsep = 8.0pt
\def\arraystretch{1.3}
\begin{table}[]
\caption{The resolutions for the frequency derivatives at $T=4$ and 10 yr (respectively from Eqs. (\ref{2212})-(\ref{232}) and Eqs. (\ref{101})-(\ref{104})). }
\begin{tabular}{|c|c|c|c|}
\hline
$T$               & 4yr & 10yr & ratio \\ \hline
$\Delta {\dot f}$ & $ 9.4\times 10^{-17} {\rm Hz\,s^{-1}}$   &   $ 9.5\times 10^{-18} {\rm Hz\,s^{-1}}$   &   0.10    \\ \hline
$\Delta {\ddot f}$ & $2.5\times 10^{-24}{\rm Hz\,s^{-2}}   $ & $ 4.6\times 10^{-25} {\rm Hz\,s^{-2}}$    &     0.18  \\ \hline
$\Delta {\dddot f}$ &   $3.2\times 10^{-31}{\rm Hz\,s^{-3}}$  &  $5.2\times 10^{-33} {\rm Hz\,s^{-3}}  $  &  0.016     \\ \hline
$\Delta {\ddddot f}$ &     &   $3.3\times 10^{-40} {\rm Hz\,s^{-4}}$   &       \\ \hline
\end{tabular}
\end{table}
\endgroup

For the third-order   phase function $\Phi_3(t)$, from  Eq. (\ref{v3}),  we have the estimation errors 
\begin{eqnarray}
\Delta {\dot f}=9.5\times 10^{-18} {\rm Hz\,s^{-1}}\\
\Delta {\ddot f}=1.0\times 10^{-25} {\rm Hz\,s^{-2}}\\
\Delta {\dddot f}=5.2\times 10^{-33} {\rm Hz\,s^{-3}}
%\Delta {\ddddot f}=5.1\times 10^{-40}
\end{eqnarray}
with 
$\Delta {\dddot f}\ll {\dddot f}$ for the highest derivative.  Similarly, from  Eq.  (\ref{v4}) for the function $\Phi_4(t)$ we obtain 
\begin{eqnarray}
\Delta {\dot f}=9.5\times 10^{-18} {\rm Hz\,s^{-1}}\label{101}\\
\Delta {\ddot f}=4.6\times 10^{-25} {\rm Hz\,s^{-2}}\label{102}\\
\Delta {\dddot f}=5.2\times 10^{-33} {\rm Hz\,s^{-3}}\label{103}\\
\Delta {\ddddot f}=3.3\times 10^{-40} {\rm Hz\,s^{-4}}\label{104}
\end{eqnarray}
now with {the desired} relation 
$ {\ddddot f}\lsim \Delta {\ddddot f}$.  Below, we use the estimation errors in Eq. (\ref{101})-(\ref{104}).
{In Table II, we compare these results with those estimated for $T=4$yr.  Because of the block diagonal structure, we can significantly improve the resolution $\Delta {\dddot f}$ with the full scaling relation $0.4^{9/2}$ mentioned before. 
}

The error $\Delta {\ddot f}$ for $T=10$yr  is 5 times smaller than that for $T=4$yr.   We then have
\begin{eqnarray}
\Delta  C_2\sim \frac{\Delta {\ddot f}}{\ddot f} \sim 1.6\times 10^{-3}, \label{dc2}
\end{eqnarray}  
and the PN effects are  likely to be still  unresolvable.   However,  for  a larger chirp mass $\mch$ and 
 a higher  frequency  $f$,  the PN  effects should be taken into account.

From Eqs. (\ref{dddp}) and (\ref{103}), we can evaluate   the estimation error for $C_3$ 
\begin{eqnarray}
\Delta  C_3\sim\frac{\Delta {\dddot f}}{\dddot f}\sim 2.4\times 10^{-2}. \label{det3}
\end{eqnarray}
%which is comparable to $\Delta C_2$  for $T=4$yr. 
As expected, we have $\Delta C_3\gg \Delta C_2 $.

\subsection{Tertiary Signatures}

Now, based on the results for $T=10$yr, we discuss the search for the tertiary effect.
For a nearly \add{circular} BBH perturbed by the reference tertiary model in Sec. II C,   we  have 
\begin{equation}
C_2-1\simeq \frac{{\ddot f}_s}{{\ddot f}_r}\simeq -6.0\times 10^{-3}\sin \psi_s,
\end{equation}
which becomes incompatible with  an isolated BBH for \add{$\sin \psi_s<0$}. 
The \add{excess}  can be measured with  the resolution $\Delta C_2$ given in Eq. (\ref{dc2}), and we can detect it at $3.6\sigma$ for the optimal phase \add{ $\sin \psi_s=-1$ (corresponding to ${\ddot f}_s\sim 2\times 10^{-24}{\rm Hz\,s^{-2}}$)}. 
For $T=4$yr, the same \add{excess} will be  totally masked by the measurement noise with  $|C_2-1|\ll \Delta C_2$.

Next, for an eccentric binary, we can examine the inconsistency between $C_2$ and $C_3$ induced by the tertiary perturbation.  In fact,  given the condition $\Delta C_3\gg \Delta C_2$,  it will be more straightforward to compare $C_3$ and  the corresponding value $K_3[K_2^{-1}(C_2)]$.   The uncertainty $\Delta C_3$ dominates the statistical error for the gap between them. From Eqs. (\ref{rr1}) and (\ref{rr3}),  the gap is approximately estimated to be  
\begin{equation}
C_3-K_3[K_2^{-1}(C_2)]\simeq \frac{{\dddot f}_s}{{\dddot f}_r}\simeq -2.6\times 10^{-2}\cos\psi_s.
\end{equation}
Given the uncertainty (\ref{det3}),
the third derivative ${\dddot f}_s$ (due to the reference  tertiary perturbation) can be detected only at $1\sigma$ level, even for the optimal orbital phase $|\cos\psi_s|=1$.

\section{discussions}
Here are some additional remarks regarding our current investigation.

\subsection{Other Harmonic Modes}
So far, we have mainly discussed observation for GW around a  single frequency $f$ (twice the orbital frequency). 
An eccentric binary emits GWs at integer times the  orbital frequency. Up  to a moderate eccentricity $e\lsim0.3$,  the third harmonic mode at the frequency $3f/2$ will be the secondary target, and   its amplitude is proportional  to $e$ with its SNR  $\sim \rho e$ ($\rho$: the  SNR of the primary mode).    As argued in  \cite{Seto:2016wom,Seto:2001pg},  by closely analyzing  the signatures  induced by the 1PN apsidal precession on the third harmonics, we might determine the total mass $M_T$ of the BBH.  

For our fiducial BBH ($\rho\sim 10$ and $f\sim 5$mHz), the third harmonics  is detectable for $e\sim 0.3$, and we can estimate the orbital eccentricity $e$ from the amplitude of the mode.  Unfortunately, the resultant error $\Delta e$  will be much larger than that from $C_2$ as discussed in Eq. (\ref{ere}).  However,  in the lower frequency regime e.g. $f\sim 1$ mHz, we cannot measure  the second derivative $\ddot f$ (see Fig. 4), and the third harmonics could be a relatively better target.

\subsection{Long-term Noise Variation}
For the block diagonalization associated with the midpoint expansion, we have assumed that the effective noise spectrum $S_n(f)$ does not depend on time (more preciously, during the course of the matched filtering analysis).  In practice, detectors may degrade over time, and we might also operate a network consisting of multiple detectors (e.g., LISA+Taiji \cite{Cai:2023ywp}) functioning at different time intervals. In such scenarios, we cannot straightforwardly apply the odd-even cancellation (\ref{cancel}) based solely on the stationarity of the noise $S_n(f)$. However, we can still adjust the time origin $t=0$ for the perturbative phase expansion and achieve similar (but more restricted)  features to the structure of the Fisher  matrix.

\section{summary}
LISA has a potential to detect stellar mass BBHs, which are similar to those already observed by ground-based detectors.  According to  a recent statistical study \cite{Seto:2022xmh}, the majority of the detected BBHs will emit nearly monochromatic GWs, exhibiting slight frequency variations within the nominal operational period of  $T\sim 4$ years. To characterize such signals, we can efficiently employ a Taylor expansion for the evolution of the GW frequency $f$. The associated derivative coefficients, such as ${\dot f}$ and ${\ddot f}$, may harbor intriguing astrophysical information, such as orbital eccentricity and tertiary perturbation effects. To facilitate this analysis, we introduced non-dimensional measures $C_n$  ($n=2,3,\cdots$) with  $C_n=1$ for an isolated circular binary. 

We  provided simple analytical expressions for the parameter estimation errors for the derivatives $f^{(n)}$,  applying the Fisher matrix formulation to a simplified phase model $\Phi_n(t)$. Then, we concretely discuss the measurability of the aforementioned astrophysical information for our fiducial BBH (chirp mass $\mch\sim30\so$, the GW frequency $f\sim 5$mHz and the signal-to-noise ratio $\rho=10$  for  $T=4$yr).  We found that  LISA can make a resolution $\Delta C_2\sim   \Delta{\ddot f}/{\ddot f}\sim 10^{-2}$, corresponding to the eccentricity error \add{$\Delta e\sim 1.3\times 10^{-2}$ for $e=0.1$}. 
For an extended operation period of $T\sim 10$yr, we might also make the next order combination $C_3$ and detect  a tertiary perturbation as small as \add{${\ddot f}_s\sim 2\times 10^{-24}{\rm Hz\,s^{-2}}$} for a nearly circular BBH. 

\add{
\begin{acknowledgments}
Equation (\ref{de}) plays an important role in this paper.  The author is deeply grateful to  Tomoya Suzuguchi for pointing out errors in its previous version.
\end{acknowledgments}
}

\bibliographystyle{mn2e}

\end{document}